\documentclass[epj]{svjour}
\usepackage{graphics,graphicx,amsmath,amssymb}
\begin{document}
\title{Nuclear Giant Resonances and Linear Response}
\author{P.-G. Reinhard\inst{1} \and Lu Guo\inst{2} \and J.A. Maruhn\inst{2} 
}                     
%
%
\institute{
Institut f\"ur Theoretische Physik II, Universit\"at Erlangen,
D-91058 Erlangen, Germany 
\and 
Institut f\"ur Theoretische Physik, Universit\"at Frankfurt,
D-60325 Frankfurt, Germany
}
\date{Received: date / Revised version: date}
%
\abstract{
We search for nonlinear effects in nuclear giant resonances (GRs), in
particular the isovector dipole and the isoscalar quadrupole modes. To
that end, we employ a spectral analysis of time-dependent Hartree-Fock
(TDHF) dynamics using Skyrme forces. Based on TDHF calculations over a
wide range of excitation amplitudes, we explore the collectivity and
degree of harmonic motion in these modes.  Both GR modes turn out to
be highly harmonic in heavy nuclei from A=100 on. There is no trace of
a transition to irregular motion and multiple resonances are
predicted. Slight anharmonicities are seen for light nuclei,
particularly for $^{16}$O. These are mainly caused by the spin-orbit
splitting.
} 
\maketitle
\section{Introduction}

Nuclei display a variety of collective excitation modes, rotation,
low-energy quadrupole or octupole vibrations, and giant resonances
\cite{Row70aB,Rin80aB}. They share this feature with many other finite
Fermion systems \cite{Ber94aB}. Nuclei provide a particularly rich
choice because they consist of two species of Fermions, protons as
well as neutrons, and are fully self-interacting Fermionic systems
without coupling to external fields (in contrast, e.~g., to ions in
metal clusters or the trap potential in quantum condensates). The low-energy
modes in the MeV range are typically large amplitude modes with
fission as an extreme example. Giant resonances (GRs), on the other
hand, are considered to oscillate with small amplitudes, although they
are often very collective to the extent that they exhaust a huge
fraction of the corresponding multipole sum rule. In that respect,
they are the dominating modes in the nuclear excitation spectrum and
have been intensively investigated over the past decades, for reviews
see, e.~g. \cite{Wou91aER,Spe91b}. The present paper concentrates on GRs
and tries to evaluate from a theoretical perspective the validity of
the assumption of small amplitude motion for them.

GRs are usually described by the well-known
Random-Phase-Approximation (RPA) \cite{Row70aB,Rin80aB}, which can be
derived as the small-amplitude limit of time-dependent Hartree-Fock
(TDHF) \cite{Row70aB,Bro71aB}. Most calculational schemes work out the
linearization of TDHF explicitly, which leads to large reductions in
computational expense for applications in restricted symmetries, see,
e.~g. \cite{Rei92a,Rei92b}. For less symmetric systems, it is
competitive and conceptually simpler to employ full TDHF and to run it
practically in the linear regime by using only small excitations. That
is the method of choice in molecules and clusters \cite{Yab96,Cal97b}
but has also been used successfully in deformed nuclei
\cite{Mar05a}. In any case, the system is analyzed in terms of its
spectrum of vibrational frequencies. That is, in fact, a classical
concept. TDHF generally provides a classical description for the
dynamics of a many-body system. Semi-classical re-quantization is
needed to recover quantized excitation states from TDHF
\cite{Kle79a,Goe82b,Neg82aR}. That is trivially and tacitly performed
in RPA: small-amplitude motion is harmonic motion and the quantization
of the harmonic oscillator identifies the vibrational frequency
$\omega$ of a mode with its excitation energy as
$E_{\rm{exc}}=\hbar\omega$.  The validity of that interpretation
depends on the ``harmonicity'' of the mode. Truly harmonic motion is
distinguished by the fact that the vibrational frequency is
independent of the amplitude. It is the aim of this paper to check
that assumption for nuclear GRs, in particular the isovector dipole GR
and the isoscalar quadrupole GR.

``Beyond RPA'' effects have been studied before from
different perspectives. Widely used is the hierarchical expansion in
$n$-particle-$n$-hole ($n$ph) states. RPA emerges in lowest order at
the level of $1$ph (and reversely $1$hp) states \cite{Rin80aB}.
The next step is to include $2$ph states, often called second RPA
\cite{Yan83a,Len90a,Lau90a}.
The coupling to the swarm of $2$ph configurations has two effects: 
a shift of the resonance position and a damping
width, often called the collisional width \cite{Ber83aR}.  The real
shift of the peak and the width are typically of the same order of
magnitude \cite{Gue93a}. The mode remains nicely harmonic if shift and
width are small, while larger values indicate nonlinear effects
contributing to the observed peak energy. An advanced analysis of the
extremely dense $2$ph spectra even allows looking for transitions from
regular to irregular motion in the nonlinear regime \cite{Lac00a}.
The extension to the $2$ph space soon grows extremely expensive,
however, and is thus often simplified by using rather small expansion
spaces \cite{Len90a,Lau90a}, schematic interactions \cite{Rei86c}, or
pre-diagonalization by selecting a subset of coupled collective states
\cite{Bor81a,Hee96a,Col01a}.

An alternative access to anharmonicities is provided by TDHF. One
merely has to watch the change of frequency of a mode with increasing
amplitude \cite{Aba92a} and in a later stage the onset of chaotic
patterns \cite{Uma85a}.  Ideally, one should use periodic TDHF orbits
to allow for a semi-classical quantization of TDHF states by computing
the classical action along one closed orbit as a function of the amplitude
\cite{Kle79a,Goe82b,Kla95a}. A quantal eigenstate is encountered
whenever the action crosses a value $(n+\frac{1}{2})\hbar$ where $n$
is some integer. Strategies for finding branches of periodic
trajectories have been successfully developed \cite{Wu97a}. They are,
however, very expensive to use for non-spherical modes and
provide far too high a spectral resolution for our purposes. We are
interested in a global test of harmonicity by watching the evolution
of the average resonance peak with increasing amplitude.  This is to
some extent along the lines of \cite{Ber97a} where a fluid-dynamical
approach to TDHF was used for that purpose.  Here, we are going to
use fully fledged TDHF without any symmetry restrictions or
constraints and invoke the systematic techniques of spectral
analysis \cite{Yab96,Cal97b,Mar05a} carried forth into the nonlinear
regime. This is an efficient method and allows to reveal the key
features of a collective mode. That has been demonstrated, e.g., in a
similar study of monopole resonances using the time-dependent
relativistic mean-field model \cite{Vre99a}.

The results of that conceptually simple study will provide insights in
several respects. First, we obtain an estimate for the excitation energy
of the quantized GR state. The difference between the excitation energy
and the estimate from the RPA frequency corresponds to the peak shift as
obtained in the elaborate $2$ph models. 
Second, we can preview the existence of double resonances for a given
nucleus. Double resonances represent second excited states of 
harmonic oscillations and are conceivable for nicely collective and
harmonic modes \cite{Lau90a,Kla95a,Bri55}.  There is, in fact,
experimental evidence for a double dipole resonance in heavy nuclei
\cite{Bor96a,Aum98aR}. Harmonicity when observed in the TDHF analysis
at varying amplitude is a strong indicator for the existence of
double, and higher, resonances.
Third and finally, one may observe a transition to large dissipation and
chaotic motion when going more deeply into the nonlinear regime
\cite{Uma85a,Vre99a}.

\section{Formal framework}

We use the techniques of \cite{Mar05a}. The TDHF equations are derived
from the Skyrme energy functional including all terms, time-even as
well as time-odd \cite{Eng75a}.  It has been worked out in recent
studies of TDHF in the large amplitude regime that a careful and
consistent management of all terms, in particular including the
time-odd contributions to the spin-orbit potential, is crucial to
avoid artefacts \cite{Mar06b}.
Wavefunctions and fields are represented on a three-dimensional
Cartesian grid in coordinate space. The fast Fourier transformation is
used to compute the action of the kinetic energy operator and to solve
the Coulomb problem.  The representation imposes no symmetry
restriction whatsoever.  This allows to explore the full phase space,
particularly in a possibly nonlinear regime. A transition to chaotic
motion would then be signified by dynamical symmetry breaking and
widespread coupling to all conceivable modes of the system (see, e.g.,
\cite{Hin93}).  The freedom is provided by our treatment but it has
yet to be seen whether the system decides to go that chaotic path.
The static mean-field equations are solved using an accelerated
gradient iteration \cite{Blu92a}. The ground state wavefunction is
excited by an instantaneous initial boost applied in the same
manner to each single-nucleon state, i.e.
$\varphi_\alpha^{\rm(gs)}
 \longrightarrow
 \varphi_\alpha=\exp{(\mathrm{i}b\hat{D})}\varphi_\alpha^{\rm(gs)}
$
where $\hat{D}$ is the multipole operator associated with the 
mode of interest and $b$ is the boost amplitude which will
translate into the amplitude of the mode with an excitation
energy $E^*\propto b^2$. The propagation of the TDHF equations is
performed with exponential expansion of the time-evolution operator
$\exp{(-\mathrm{i}\hat{h}_{\mathrm{mf}}t)}$
\cite{Flo78a}. The expectation value of the multipole momentum 
$\overline{D}(t)=
\sum_\alpha\left(\varphi_\alpha(t)|\hat{D}|\varphi_\alpha(t)\right)
$
is recorded along the dynamical evolution. Finally, the signal
$\overline{D}(t)$ is Fourier transformed into the frequency domain to
obtain the spectral distribution of multipole strength. Some
filtering is required to avoid artefacts in the spectra. We use
windowing in the time domain with $\cos^n{(\pi t/2T_{\rm fin})}$ and a
large $n$ to have a short effective time window with width of about
250 fm/c. This is chosen to smoothen spectral details and to
concentrate the analysis on the average resonance peak.  Such a strong
filtering allows also to use a conveniently sized numerical box
\cite{Rei06a}. We actually employ a 24$\times$24$\times$24 grid with
spacing of 1 fm.

There exists a great variety of Skyrme parametrizations which perform
equally well concerning bulk properties of the ground state but differ
in details for more refined observables \cite{Ben03aR}. We find that
all parametrizations which provide a reasonable description of GRs in
heavy nuclei perform very similarly concerning (an-)harmonicity. Thus we
confine the presentation here to the single parametrization SLy6 
from  \cite{Cha98a}.

As test cases,we chose a selection of doubly magic nuclei from
$^{16}$O up to $^{208}$Pb. Results for the isovector dipole GR are
shown over the whole sample. The trends turn out to be very similar
for the isoscalar quadrupole GR and the discussion is reduced to only
two typical nuclei.

\section{Results and discussion}

\begin{figure*}
\centerline{
\includegraphics[width=176mm]{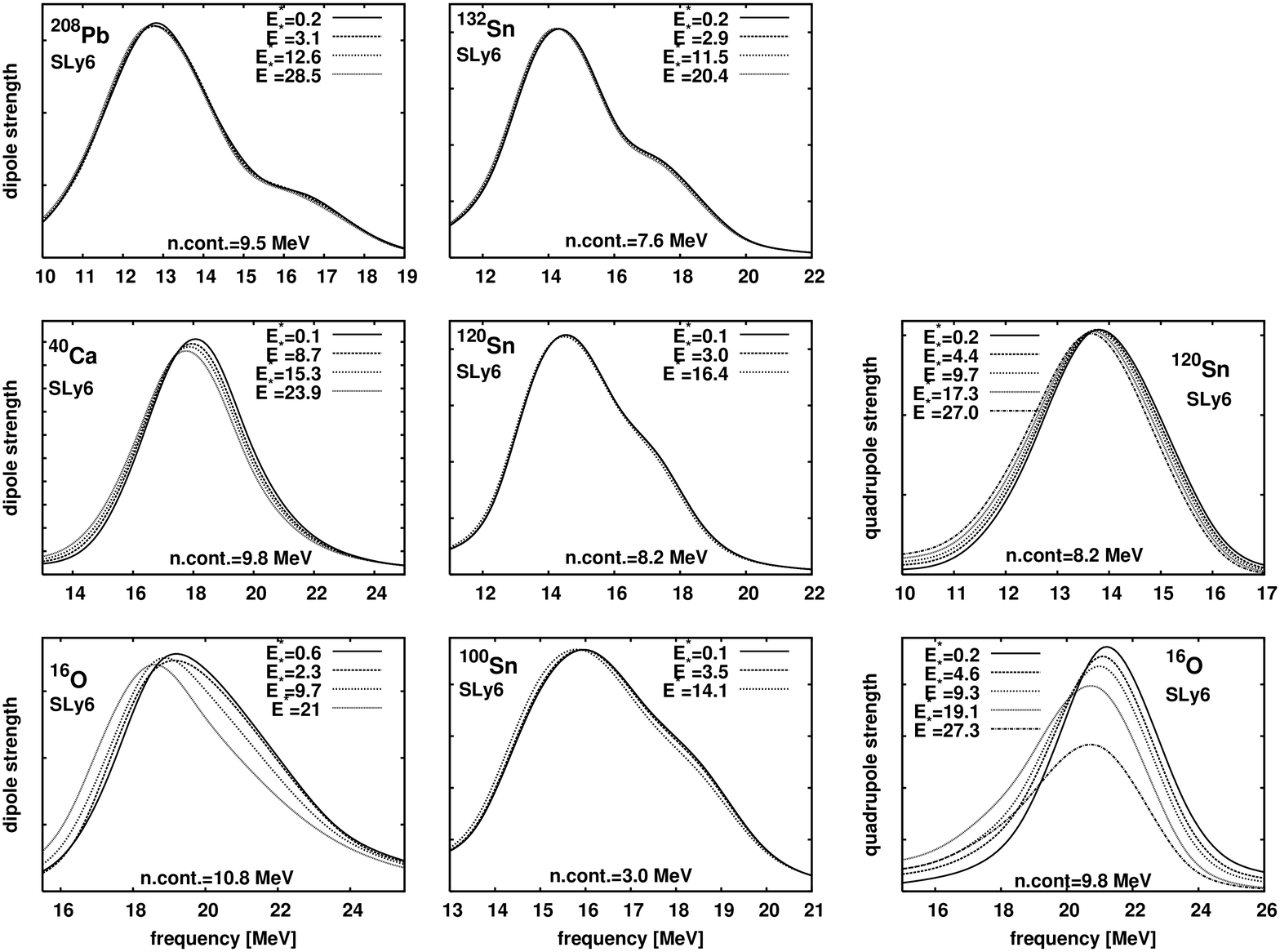}
}
\caption[zu]{\label{fig:spectr-all} 
Strength distributions for isovector dipole (left six panels)
and isoscalar quadrupole (rightmost two panels) 
GR modes in six spherical
nuclei. The modes were excited with an instantaneous boost at
different boost energies as indicated. All calculations were done
with the Skyrme parametrization SLy6. The threshold for neutron
emission into the continuum, ``n.cont.'', is indicated in  the plots.
}
\end{figure*}

Figure \ref{fig:spectr-all} shows the spectral distributions with
varying excitation amplitudes for the test cases in our sample.  Note
the large width which was deliberately chosen to override the details
of spectral fragmentation over the $1$ph states in the vicinity of the
GR (the analogue of Landau damping in bulk matter).  That
fragmentation width varies from 4 MeV in $^{16}$O down to 2 MeV for
$^{208}$Pb. For spectral filtering, we use a fixed folding width
of 1.6 MeV for all nuclei, which suffices to suppress the details and
always deliver a unique and well-localized resonance. The plots show a
narrow frequency window about the resonance in order to better
visualize the drift of the peaks with excitation amplitude. Other
parts of the spectrum can safely be ignored, because there is not much
strength outside the shown window. Within the plotting window, we
see nicely concentrated resonance peaks in all cases. That fact alone
already carries a message: the isovector dipole and isoscalar
quadrupole resonances stay well inside the regime of regular motion
for all amplitudes considered here, and probably far beyond that
point. That agrees with the findings of \cite{Ber97a}. Note that the
isovector monopole mode was found to be closer to the chaotic regime
\cite{Vre99a} and that the highly nonlinear low-energy modes
also very quickly turn chaotic \cite{Uma85a}.

\begin{figure*}
\centerline{
\includegraphics[width=150mm]{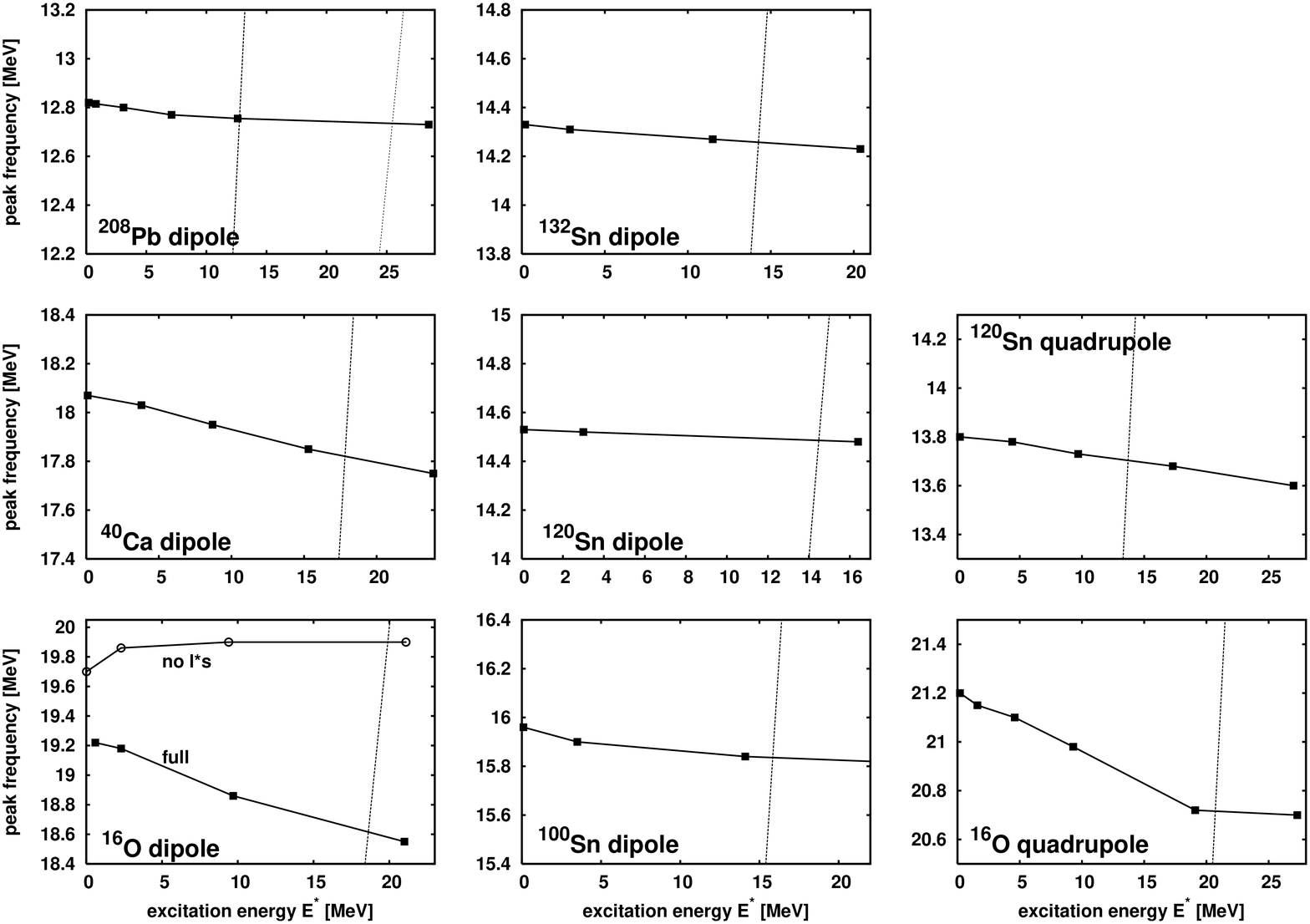}
}
\caption[zu]{\label{fig:spectr_trends} 
Trend of resonance peak frequency with excitation energy
for isovector dipole and isoscalar quadrupole resonances
in various spherical nuclei as indicated. All calculations  were
done with the Skyrme parametrization SLy6. The diagonal of
``peak-frequency = excitation-energy'', i.e.
$E_{\rm exc}=\hbar\omega$, is indicated by a faint
straight line. In case of $^{208}$Pb, we also show a second steep line
which stands for the the second resonance mode with
$E_{\rm exc}=2\hbar\omega$.
In case of $^{16}$O, results from a calculation without
spin-orbit force are indicated by open circles.
}
\end{figure*}

The feature of interest for our studies are the trends with increasing
amplitude. Reading the left three panels in figure
\ref{fig:spectr-all} from bottom up, it becomes obvious that small
nuclei show some changes while the spectra of heavy nuclei stay
extremely robust. The middle panel checks the effect along an isotopic
chain, here the Sn isotopes. Note that this represents not only
isotopic trends but also dramatic differences in overall binding with
$^{100}$Sn being at the edge of stability, $^{132}$Sn on the neutron
rich side not much better bound, and $^{120}$Sn well inside the valley
of stability.  There is very little difference in spectral pattern in
view of these dramatic variations in isospin and stability. System
size seems to be the key feature which determines the robustness of
the resonance. At second glance, one  finds that the two unstable
nuclei (upper middle and lower middle panels) indicate indeed slightly
larger drift of the resonance peaks than the stable $^{120}$Sn. But
this remains a faint effect. As a rule of thumb, we find that the
dipole resonance is a harmonic mode in nuclei with $A\geq 100$.

The right panels in figure \ref{fig:spectr-all} show two samples for
the isoscalar quadrupole mode. The changes are still small but
generally stronger than for the dipole modes in the same nuclei. Note
that the stronger shifts develop towards the lower end of the
spectrum. This is caused by the much more nonlinear low-lying
quadrupole mode \cite{Uma85a} which, although having comparatively
small strength, affects the lower side of the resonance peak.  Note
also that the resonance as such is much more robust and the drift of
peak position is not much larger than in the dipole case. Thus we can
generalize the conclusion from above and call also the isoscalar
quadrupole resonance a harmonic mode for medium-heavy and heavy
nuclei.

Figure \ref{fig:spectr_trends} summarizes the trends of peak positions
with excitation energy.  It confirms what we could read off (less
quantitatively) from the previous figure. The dipole resonance is
harmonic for $A\geq 100$.  The quadrupole still shows slightly more
drift in $^{120}$Sn, but also that can be considered as small and
is further reduced for larger nuclei. 

The figure also shows faint steep lines which indicate the
simple-minded oscillator quantization $E_{\rm exc}=\hbar\omega$.  The
point where both lines cross each other indicates the amplitude of the
oscillation which contains precisely one oscillator quantum.  It is
practically the oscillation frequency in the linear regime (extremely
small amplitudes) for the harmonic modes in $A\geq 100$.  There are
sizeable corrections for the lighter nuclei. The actual resonance peak
is 0.6 MeV below the RPA frequency for $^{16}$O and 0.25 MeV below for
$^{40}$Ca. 
For the particularly harmonic case of $^{208}$Pb, we also analyze the
occurrence of the double resonance excitation, see the second steep line
in the upper left panel of figure \ref{fig:spectr_trends}. The mode
remains safely harmonic up to the double resonance regime which
confirms once more the existence of a double resonance without any
frequency shifts \cite{Bor96a}.

It is known that the spin-orbit force has a tendency to degrade
collectivity and enhance dissipation \cite{Rei88b}. We have tested
that for our worst case of $^{16}$O. The lower left panel of figure
\ref{fig:spectr_trends} shows also results from TDHF calculations
without spin-orbit force. We see indeed more harmonicity than for the
full force. This confirms again that the spin-orbit force to some
extent counteracts collectivity in GRs. The mechanism is that the
spin-orbit splitting enhances spectral fragmentation which, in turn,
reduces the coherence in the superposition of the many $1$ph modes to
one dominant collective mode \cite{Row70aB,Rin80aB}. Without
spin-orbit force, GR would be highly harmonic modes all over the
periodic table including light nuclei.

\section{Conclusion}

We have analyzed the isovector dipole and isoscalar quadrupole
giant resonance (GR) in nuclei of different size using fully fledged
time-dependent Hartree-Fock (TDHF) dynamics in connection with Skyrme
forces.  The aim was to check the commonly accepted view that GRs are
collective modes corresponding to rather harmonic oscillations of the
nuclear many-body system. To that end, we analyzed the spectral
distributions of the corresponding multipole moments for widely
varying amplitudes of the oscillations and for a broad range of nuclei
between the rather small $^{16}$O and the heavy $^{208}$Pb.
We find in all cases that the GRs remain in the regime of regular
motion confirming from that aspect the robust collectivity of these
modes. ``Harmonicity'' is more critical. We see it extremely well
fulfilled for heavy nuclei, at least from $A=100$ on. 
This, in turn, suggests the existence and robustness of
highly collective multiple-resonance excitations which have been
experimentally assessed already for the double resonance in heavy
nuclei. 
Light nuclei show first glimpses of nonlinear effects to the extent
that the resonance-peak position shows visible drift with increasing
amplitude. For $^{16}$O, e.g., we observe a drift of 0.6 MeV from the
strictly linear regime (small amplitudes) to the realistic amplitude
where the oscillator quantization recipe
$E_{\mathrm{esc}}=\hbar\omega$ is satisfied. It is still a small, but
non-negligible, effect.  The isoscalar quadrupole mode is similarly
robust concerning the peak drift. The patterns of the spectral
distribution, however, are a bit more volatile at the lower tail,
which is due to the perturbing influence of the highly nonlinear
low-energy collective surface quadrupole mode.

\medskip

Acknowledgments:
Lu Guo acknowledges the support from Alexander von Humboldt Foundation.
We gratefully acknowledge support by the Frankfurt Center for
Scientific Computing and from the BMBF (contracts 06 ER 124 and
06 F 131).

 \bibliographystyle{prsty}
 \bibliography{nonlres}

\end{document}